\documentclass[article]{revtex4}%
\usepackage{amsfonts}
\usepackage{amsmath}
\usepackage{amssymb}
\usepackage{graphicx}%
\usepackage{subcaption}
\graphicspath{{./images/}}
\providecommand{\U}[1]{\protect\rule{.1in}{.1in}}

\begin{document}
\title{Vortex Solitons and Filamentation of Electromagnetic Beams in Relativistically
Degenerate Plasmas}
\author{N. Maltsev$^{1}$ and V.I. Berezhiani$^{1,2}$ }
\affiliation{$^{1}$School of Physics, Free university of Tbilisi, Tbilisi 0159, Georgia}
\affiliation{$^{2}$Andronikashvili Institute of Physics (TSU), Tbilisi \ 0177, Georgia.}
\affiliation{}


\begin{abstract}
We study the propagation and stability of electromagnetic vortex beams in
relativistically degenerate plasmas. We show that such plasmas support
localized vortex solitons carrying orbital angular momentum and analyze their
linear and nonlinear stability. Vortex solitons undergo azimuthal
symmetry-breaking instabilities whose growth rates depend on beam power,
propagation constant, and topological charge, with the dominant mode
determining the number of filaments formed during breakup. We further
demonstrate that vortex solitons act as nonlinear attractors with a finite
basin of attraction, while the vortex core remains topologically protected,
maintaining a strictly zero field intensity at the beam center throughout the
evolution. The results persist across a broad range of degeneracy parameters
and are relevant to hard $X$-ray radiation propagating in dense astrophysical plasmas.

\end{abstract}
\startpage{1}
\endpage{1}
\maketitle

\bigskip

\section{\bigskip Introduction}

Astrophysical systems are characterized by intense electromagnetic radiation
whose spectral content spans a broad range, from radio waves to hard $X$-ray
and $\gamma$-ray emission \cite{Begelman}. Most cosmic objects - such as
stars, interstellar and intergalactic media, active galactic nuclei, and
quasars - are either composed of plasma or embedded in a plasma environment
\cite{Zhel}. In many of these systems, radiation propagates through regions of
extremely high density, where collective plasma effects play a dominant role
in shaping wave dynamics. For the ultrahigh plasma densities relevant to
compact astrophysical objects, including the interiors of white dwarfs,
neutron stars, pre-supernova cores, and possibly the central engines of
gamma-ray bursts \cite{Shapiro}, the classical plasma description becomes
inadequate. Under such conditions, the electron gas is degenerate and must be
described using Fermi--Dirac statistics. The corresponding Fermi energy
exceeds the atomic binding energy, leading to complete ionization of matter.
Typical electron number densities in these environments are believed to lie in
the range from $10^{26}cm^{-3}$ to $10^{34}$ $cm^{-3}$, where the average
interparticle spacing becomes smaller than the thermal de Broglie wavelength
and the plasma behaves as a weakly coupled degenerate system \cite{Landau}. At
the same time, compact astrophysical systems are known to host extremely
intense high-frequency radiation fields. In particular, strong and spectrally
narrow $X$-ray emission may arise from cyclotron harmonics in strongly
magnetized environments \cite{Harding}, atomic line emission from highly
ionized species in accretion-powered sources \cite{Fabian}, or external
irradiation by localized compact objects such as neutron stars and white
dwarfs \cite{Lewin}. In many situations of interest, dense plasma layers
remain transparent to X-ray radiation due to the suppression of collisional
absorption at high frequencies and the dominance of electron degeneracy
effects, allowing electromagnetic waves to propagate over macroscopic distances.

A series of studies has examined nonlinear interactions between high-frequency
electromagnetic (EM) radiation and plasma waves in degenerate media.
Stimulated scattering processes were analyzed in Refs. \cite{Chanturia}%
-\cite{Goshadze}, demonstrating the emergence of instabilities in both weakly
and strongly degenerate electron plasmas. Refs. \cite{Tsintsadze}-\cite{Sound}
investigated solitary solutions in fully degenerate relativistic and
nonrelativistic multi-species plasmas, showing that such nonlinear structures
are stable against small perturbations and persist across the full range of
degeneracy levels.

The self-trapped propagation of EM beams in fully degenerate relativistic
electron--positron and electron plasmas has been investigated in
(1+2)-dimensional geometries \cite{ep-Nana},\cite{me-soso} demonstrating the
existence of radially symmetric, localized EM structures supported by
self-guiding mechanisms in these plasmas.

Electromagnetic beams carrying orbital angular momentum, known as vortex
beams, possess a phase singularity and an annular intensity profile. Under
laboratory conditions, vortex beams can be produced using phase plates,
holographic techniques, or mode-conversion schemes (see Ref. \cite{Shen} and
references therein). In astrophysical settings, twisted EM fields may arise
naturally due to rotation of compact objects, frame-dragging effects, or
propagation through inhomogeneous media, imprinting angular momentum onto
radiation \cite{Harwit},\cite{Tamburini}.

Recent work \cite{Zaza} examined the nonlinear propagation of
orbital-angular-momentum--carrying electromagnetic beams in highly degenerate
relativistic electron--positron plasmas. Because the high-frequency
electromagnetic pressure acts equally on electrons and positrons, charge
separation is suppressed, and, within the paraxial approximation, the beam
dynamics is governed by a nonlinear Schr\"{o}dinger equation with saturating nonlinearity.

In the present work, we base our analysis on the system of nonlinear equations
describing electromagnetic beam propagation in degenerate electron plasmas,
derived in previous studies \cite{Goshadze},\cite{me-soso}. We extend this
framework to investigate the propagation of vortex electromagnetic beams.
Unlike in the electron--positron plasma case, the system cannot be reduced to
a single nonlinear Schr\"{o}dinger equation; instead, one must analyze a
coupled set of nonlinear equations. The presence of orbital angular momentum
qualitatively modifies the nonlinear interaction. We demonstrate that
degenerate electron plasmas can support localized vortex-type electromagnetic
structures whose properties differ from those of non-vortex beams, thereby
highlighting the role of beam topology in relativistic plasma dynamics.

\section{Model}

As shown in Ref. \cite{Goshadze}, the dynamics of relativistic electromagnetic
beams propagating in a degenerate electron plasma can be described by a set of
nonlinear equations which, in dimensionless form, read as follows.%

\begin{equation}
2i\frac{\partial A}{\partial z}+\nabla_{\perp}^{2}A+\left(  1-\frac{N}{1+\Psi
}\right)  A=0 \label{1}%
\end{equation}

\begin{equation}
\nabla_{\perp}^{2}\Psi+1-N=0 \label{2}%
\end{equation}

\begin{equation}
N=\frac{\left(  1+\Psi\right)  \left[  \left(  1+\Psi\right)  ^{2}-\left(
1+\left\vert A\right\vert ^{2}-d\right)  \right]  ^{3/2}}{d^{3/2}\left[
\left(  1+\Psi\right)  ^{2}-\left\vert A\right\vert ^{2}\right]  ^{1/2}}
\label{3}%
\end{equation}

Here $A$ is the slowly varying amplitude of circularly polarized vector
potential $e\mathbf{A/}\left(  m_{e}c^{2}\Gamma_{0}\right)  \mathbf{=}\left(
\widehat{\mathbf{x}}+i\widehat{\mathbf{y}}\right)  A\exp(-i\omega_{0}%
t-k_{0}z)+c.c.$, where $\widehat{\mathbf{x}}$ and $\widehat{\mathbf{y}}$ are
unite vectors transverse to the electromagnetic beam propagation direction $z$
. The field frequency $\omega_{0}$ and wave vector $k_{0}$ satisfy the
dispersion relation $\omega_{0}^{2}=k_{0}^{2}c^{2}+\omega_{e}^{2}/\Gamma_{0}$
, with $\omega_{e}=\left(  4\pi e^{2}n_{0}/m_{e}\right)  ^{1/2}$ \ being the
plasma frequency, $n_{0}$ the equilibrium electron density, and $\Gamma
_{0}=\left(  1+R_{0}^{2}\right)  ^{1/2}$ the generalized relativistic factor,
where $R_{0}=\left(  n_{0}/n_{c}\right)  ^{1/3}$ with $n_{c}=m_{e}^{3}%
c^{3}/3\pi^{2}\hbar^{3}=5.9\times10^{29}cm^{-3}$. The normalized potential is
defined as $\Psi=e\varphi/\left(  m_{e}c^{2}\Gamma_{0}\right)  $ where
$\varphi$ is the charge-separation scalar potential generated by the action of
the high-frequency electromagnetic pressure on electrons. The normalized
electron density is $N=N/n_{0}$ . The dimensionless spatial variables are
introduced as $z=\left(  \omega_{e}^{2}/c\omega_{0}\Gamma_{0}\right)  z$, and
$\mathbf{r}_{\perp}=\left(  \omega_{e}/c\sqrt{\Gamma_{0}}\right)
\mathbf{r}_{\perp}$. In Eq. (3) the parameter $d=R_{0}^{2}/\left(  1+R_{0}%
^{2}\right)  $ quantifies the level of degeneracy: for the weakly degenerate
case $\left(  R_{0}<<1\right)  $ $d\simeq R_{0}^{2}$ , whereas in the
relativistically degenerate regime $\left(  R_{0}>>1\right)  $ $d\rightarrow1$.

Deriving the system of Eqs. (1)-(3) it is assumed that plasma is highly
transparent $\omega_{e}/\omega_{0}<<1$ and $\lambda<<L_{\perp}<<L_{\parallel}$
where $\lambda\approx2\pi c/\omega_{0}$ is the wavelength of EM radiation,
$L_{\perp}$ and $L_{\parallel}$ are the characteristic longitudinal and
transverse spatial dimensions of the EM beam. This system describes the
dynamics of strong amplitude narrow EM beams in plasma with arbitrary (but
physically justified) strength of degeneracy. The gas approximation for
degeneratye palsma implyes that particle interaction should be less the Fermi
energy. This condition implies that plasma electron density should be
$n_{0}\geq e^{6}m_{e}^{3}/\hbar^{6}=\allowbreak6.\,7\times10^{24}cm^{-3}$
\ and consequently $R_{0}>>0.02$ , note that for $n_{0}=n_{c}=5.9\times
10^{29}cm^{-3},$ $R_{0}=1$ while $n_{0}=10^{34}cm^{-3}$ corresponded to the
ultra-relativistic degeneracy with $R_{0}\approx26$. \ For the electron
densities $10^{26}cm^{-3}-10^{34}cm^{-3}$ the level of degeneracy measure $d$
in Eq.(3) varies in the range $0.003<d<0.999$.

For strongly relativistic degenerate plasma $(d\rightarrow1)$ we get the
following set of equations%
\begin{equation}
2i\frac{\partial A}{\partial z}+\nabla_{\perp}^{2}A+\left(  1+\left\vert
A\right\vert ^{2}-\left(  1+\Psi\right)  ^{2}\right)  A=0 \label{4}%
\end{equation}

\begin{equation}
\nabla_{\perp}^{2}\Psi+1+\left(  1+\Psi\right)  \left(  \left\vert
A\right\vert ^{2}-\left(  1+\Psi\right)  ^{2}\right)  =0 \label{5}%
\end{equation}

Let us now consider the solitary wave solutions carrying vortices. Assuming
that solutions in polar coordinates are of the form $A=\widehat{A}\left(
r\right)  \exp\left(  im\theta+ikz/2\right)  $ and $\Psi=\widehat{\Psi}\left(
r\right)  $ where $r=\sqrt{x^{2}+y^{2}}$ and $\theta$ is the polar angle, Eqs.
(4) and (5) reduce to ordinary differential equations for the real valued
amplitude $\widehat{A}$ and potential $\widehat{\Psi}$ as follows:%

\begin{equation}
\frac{d^{2}\widehat{A}}{dr^{2}}+\frac{1}{r}\frac{d\widehat{A}}{dr}%
-k\widehat{A}-\frac{m^{2}}{r^{2}}\widehat{A}+\left(  1+\widehat{A}^{2}-\left(
1+\widehat{\Psi}\right)  ^{2}\right)  \widehat{A}=0 \label{6}%
\end{equation}

\begin{equation}
\frac{d^{2}\widehat{\Psi}}{dr^{2}}+\frac{1}{r}\frac{d\widehat{\Psi}}%
{dr}+1+\left(  1+\widehat{\Psi}\right)  \left(  \widehat{A}^{2}-\left(
1+\widehat{\Psi}\right)  ^{2}\right)  =0 \label{7}%
\end{equation}
Here $k$ is a propagation constant and $m(\neq0)$ is an integer known as the
topological charge of the vortex.

We have used numerical methods to determine localized solutions of Eqs.
(6)--(7) subject to the boundary conditions $\left(  \widehat{A}%
,\widehat{\Psi}\right)  \rightarrow0$ as $r\rightarrow\infty$ , while
$\widehat{A}_{r\rightarrow0}\rightarrow A_{0}r^{\left\vert m\right\vert }$ ,
$\widehat{\Psi}_{r\rightarrow0}\rightarrow\Psi_{0}$, $\left(  d\Psi/dr\right)
_{r\rightarrow0}\rightarrow0$. In what follows, we consider only the
lowest-order (lowest radial eigenmode) solutions of Eqs. (6) and (7). Such
solutions imply that the vector potential has a node at the origin $r=0$
reaches a maximum at a finite radius, and then monotonically decreases with
increasing $r$, whereas the scalar potential is nodeless, also reaches a
maximum, and decays with increasing $r$. To obtain such solutions for a given
propagation constant $k>0$, one must determine the appropriate values of the
constants $A_{0}$ and $\Psi_{0}$ with high accuracy. This is achieved
numerically by solving Eqs. (6)--(7) as a boundary-value problem, employing a
shooting method in which the parameters $A_{0}$ and $\Psi_{0}$ are iteratively
adjusted to satisfy the localization conditions at large radii. \ For $m=1$
and $k=0.2$, localized solitary solutions are found for $A_{0}=0.368$ and
$\Psi_{0}=0.065$. The corresponding radial profiles of the electromagnetic
fields and the electron density are shown in Fig. 1(a). Similar field
structures are obtained for other values of the propagation constant provided
that $k<1$. In Fig. 1(b), we present the field and
density profiles for $k=0.3$. It is evident that, with increasing $k$ , the
amplitudes of the field components $\widehat{A}_{m}$ and $\widehat{\Psi}_{m}$
increase. As shown in Fig. 2, these amplitudes are monotonically increasing
functions of $k$. We emphasize that, simultaneously, the electron density
becomes increasingly depleted as the potentials reach their maximal values at $r_{m}$, and in the limit $k\rightarrow
k_{c}=0.3$ the density tends to zero, indicating the onset of electron cavitation.
Note, however, that for $k>k_{c}$ the electron density becomes negative, and
therefore the plasma model based on a fluid description is no longer applicable.

\begin{figure}[h]
    \centering
        \includegraphics[width=0.4\linewidth]{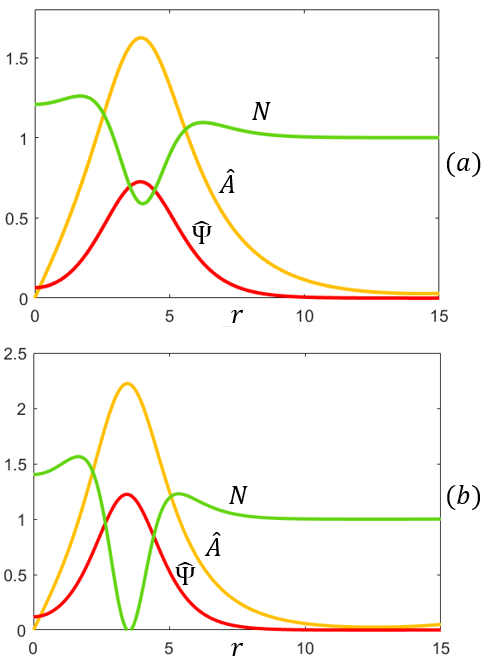}
        \caption{Numerically found stationary potential profiles for $m=1$, $k=0.2$ (a) and $k=0.3$ (b)}
    \hfill
    \label{fig 1}
\end{figure}

\begin{figure}[h]
    \centering
        \includegraphics[width=0.45\linewidth]{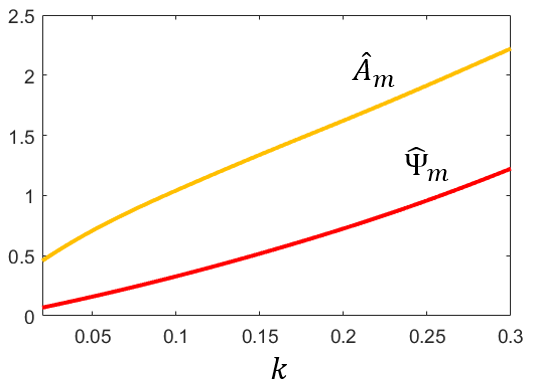}
        \caption{Stationary $m=1$ vortex amplitudes as functions of $k$}
    \hfill
    \label{fig 2}
\end{figure}

A similar qualitative behavior is observed for vortex solutions with higher
topological charge. In particular, for $m=2$, localized solutions also exist
provided that the propagation constant satisfies the same condition as in the $m=1$ case. Within this domain, higher-charge vortices exhibit the
same generic features as the fundamental vortex, including annular field
localization and electron density depletion leading to cavitation as
$k\rightarrow k_{c}$. Not only does changing the topological charge preserve the existance domain, but it also causes negligible changes to the cavitation condition and the amplitude dependence on the propogation constant. As shown of Fig. 3, the vortex solution for $k=0.2$ and $m=2$ is just a wider version of its lower topological charge predecessor.

\begin{figure}[h]
    \centering
        \includegraphics[width=0.5\linewidth]{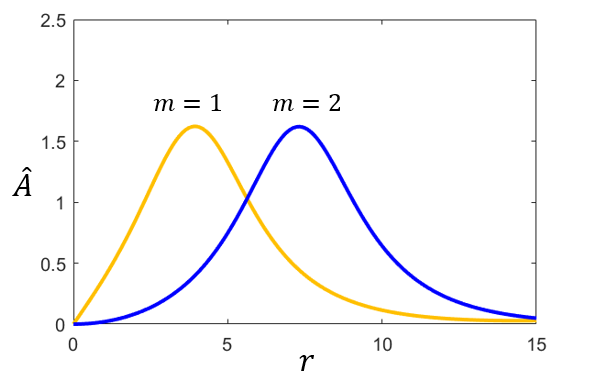}
        \caption{Numerically found stationary potential profiles for $k=0.2$, $m=1$ and $m=2$}
    \hfill
    \label{fig 3}
\end{figure}

The EM beam power $P=\int d\mathbf{r}_{\perp}\left\vert A\right\vert ^{2}$ is
an integral of motion of the system of Eqs. (1)--(3). For an EM field carrying
a nonzero topological charge, the system admits an additional conserved
quantity, namely the angular momentum,%

\begin{equation}
L=\frac{i}{2}\int d\mathbf{r}_{\perp}\left[  x\left(  A^{\ast}\frac{\partial
A}{\partial y}-c.c.\right)  -y\left(  A^{\ast}\frac{\partial A}{\partial
x}-c.c.\right)  \right]  \label{8}%
\end{equation}
This expression represents the paraxial approximation to the orbital angular
momentum of the EM field. One can readily show that, for steady-state
solutions, the angular momentum is fully determined by the topological charge
and the beam power, yielding the relation $L=mP$.

As follows from numerical simulations of Eqs. (6) and (7), the power trapped
in self-guided vortex solitons,%

\begin{equation}
P=2\pi\int_{0}^{\infty}drr\widehat{A}^{2} \label{9}%
\end{equation}
is a monotonically increasing function of the propagation constant $k$
$\left(  dP/dk>0\right)  $. Since $k$ is itself an increasing function of
$\widehat{A}_{m}$, it is convenient to present the dependence of the trapped
power on $\widehat{A}_{m}$, as shown in Fig. 4. One can see that, in the
strongly degenerate case, the power trapped in a singly charged ($m=1$) vortex
solitonic structure is bounded from below by a finite value $P_{0}=145$, which
is reached at small amplitudes $\widehat{A}_{m}<<1$ (i.e., in the limit
$k\rightarrow k_{c}$). The upper bound of the trapped power, $P_{c}=283$, is
associated with electron cavitation occurring at larger amplitudes
$\widehat{A}_{mc}=2.221$ (corresponding to $k\rightarrow k_{c}$).
\begin{figure}[h]
    \centering
        \includegraphics[width=0.45\linewidth]{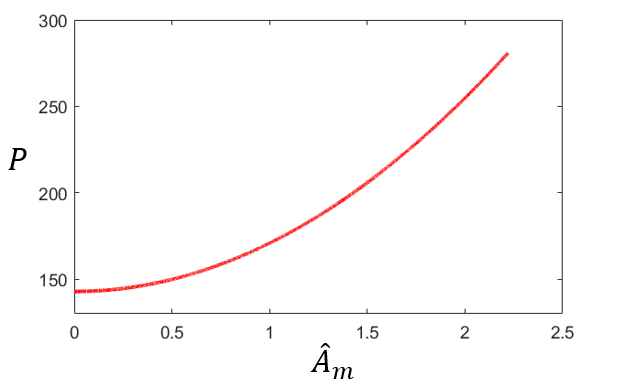}
        \caption{Power carried by the stationary state as a function of its amplitude}
    \hfill
    \label{fig 4}
\end{figure}

We now turn to the stability analysis of vortex solitons. Numerous analytical
and numerical studies have shown that ring-shaped vortex beams described by
nonlinear Schr\"{o}dinger equations with saturating nonlinearity in various
media are susceptible to azimuthal symmetry-breaking modulational instability
\cite{Skryabin}, \cite{Kivshar}. Although the present system differs from the
standard nonlinear Schr\"{o}dinger equation due to the coupling between the
electromagnetic field and the plasma response, the qualitative features
responsible for the azimuthal instability remain similar. In particular, the
self-focusing nonlinearity combined with the ring-shaped intensity
distribution of vortex solitons creates favorable conditions for azimuthal
modulation growth. We therefore expect that vortex solitons supported by
degenerate relativistic electron plasmas will exhibit a qualitatively similar
instability scenario. The stability of vortex solitons can be analyzed by
following the linear stability procedure developed in \cite{Atai} (see also
\cite{Berge}). In this approach, one considers small perturbations acting
along a ring of mean radius $r_{m}$ where $\widehat{A}(r_{m})=\widehat{A}_{m}$
and $\widehat{\Psi}\left(  r_{m}\right)  =\widehat{\Psi}_{m}$. Assuming
constant intensities and spatial uniformity for this ring, one can rewrite the
operator in (6) and (7) as $\nabla_{\perp}^{2}=r_{m}^{-2}\partial^{2}%
/\partial\theta^{2}$. The growth rate of azimuthal perturbations with phase
factor $\phi=Kz+M\theta$, where $M$ is an integer azimuthal mode number, can
then be derived in the form%

\begin{equation}
\operatorname{Im}(K)=\frac{1}{2}\frac{M}{r_{m}}\left[    2\widehat{A}_m^2-\frac{M^{2}}{r_{m}^{2}}-Q\right]  ^{1/2}.
\label{10}%
\end{equation}
Where
\begin{equation}
Q=4\widehat{A}_m^2\left(1+\widehat{\Psi}_m\right)^2\left({\frac{M^{2}}{r_{m}^{2}}+\frac{1}{1+\widehat{\Psi}_m} +2\left(1+\widehat{\Psi}_m\right)^2}\right)^{-1}
\end{equation}
The function under the root is positive for some integer values of $M$ in the studied range of propagation constant $k$, implying the the existence of azimuthal instabilities.\ In
Fig. 5 we plot $\operatorname{Im}(K)$ versus $M$ for different values of $k$. The parameters
$r_{m}$, $\widehat{A}_{m}$ and $\widehat{\Psi}_{m}$ are all calculated via the localized solutions of equations (6) and (7). The
integer value of $M$ for which $\operatorname{Im}(K)$ is maximal gives the
approximate number of modulations that affect the ring.

\begin{figure}[h]
    \centering
        \includegraphics[width=0.6\linewidth]{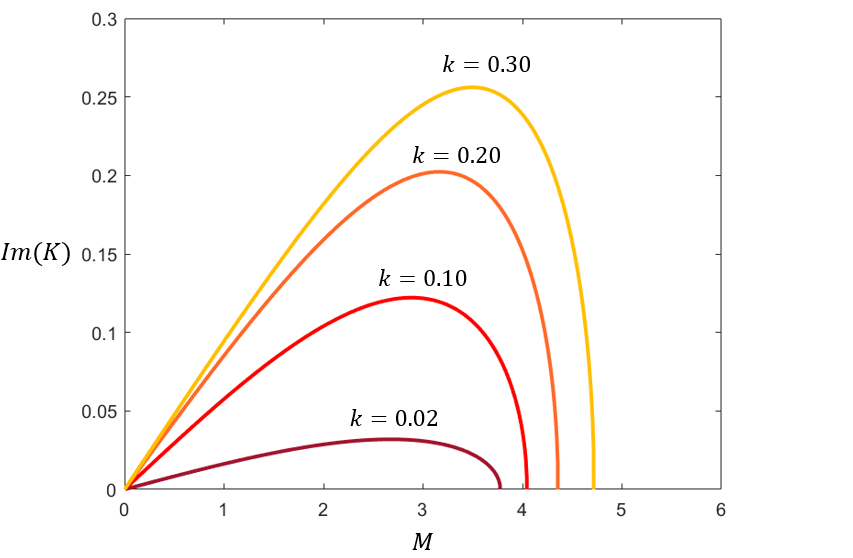}
        \caption{$\operatorname{Im}(K)$ as a function of $M$ for $k=0.02$, $ 0.10$, $0.20$ and $0.3$}
    \hfill
    \label{fig 5}
\end{figure}

To verify the predictions of the linear stability analysis, we numerically
solved the system of Eqs. (4) and (5) using finite-difference methods. As
initial conditions, we employed the stationary solutions of Eqs. (6)--(7)
obtained for different values of the propagation constant $k<k_{c}$. Since any
finite-difference discretization approximates partial differential equations
with limited accuracy, the use of stationary solutions as initial conditions
introduces an inherent small perturbation. Our numerical simulations show that
this perturbation rapidly triggers an instability, in agreement with the
predictions of the linear stability analysis.

In Fig. 6, we present the results of a representative simulation in which the
initial condition corresponds to $k=0.2$. We split each discrete step into linear and non-linear parts, respectively using Crank-Nicolson and Euler methods for their solution. One observes that the instability
leads to the breakup of the vortex soliton into four filaments, in accordance
with the dominant azimuthal mode predicted by the linear stability analysis.
The breakup occurs at $z_{br}\approx70$ while the corresponding diffraction
length is $z_{D}\approx20$.
This implies that the vortex solitary wave decays after propagating over $3.5$
diffraction lengths. As shown in Fig. 7 the filaments propagate tangentially away from the original ring-shaped
intensity distribution, conserving the total angular momentum $L$ of the
electromagnetic field. As they separate, the filaments evolve into stable,
spatially localized solitonic beams carrying zero topological charge. Similar
behavior is observed for vortex solitons with higher values of $m$.
\begin{figure}[h]
        \centering
        \includegraphics[width=0.4\linewidth]{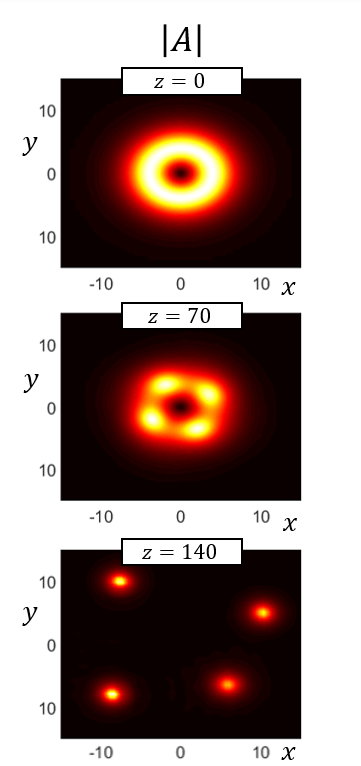}
        \caption{Numerically achieved evolution of $k=0.2$, $m=1$ vortex soliton}
        \label{fig 6}
\end{figure}

\begin{figure}[h]
        \centering
        \includegraphics[width=0.4\linewidth]{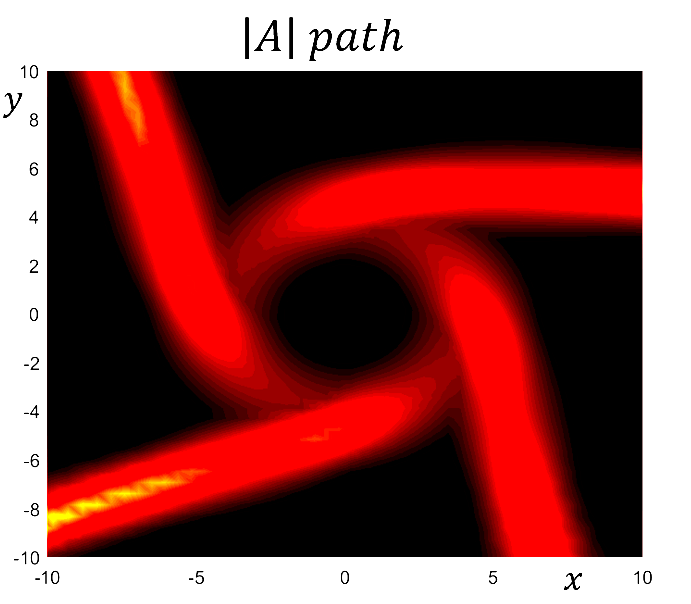}
        \caption{The soliton trajectories for the $k=0.2$, $m=1$ vortex}
        \label{fig 7}
\end{figure}

The next question to investigate concerns the nonlinear stability of
electromagnetic fields carrying topological charge under non-ideal excitation
conditions, when the initial spatial profile and field strength differ
substantially from those of the stationary vortex soliton solution. This
analysis is particularly relevant for realistic beam generation, where exact
soliton profiles cannot be prepared. For this purpose, we consider a Gaussian
input beam of the form%

\begin{equation}
A(z=0,x,y)=\left(  x+iy\right)  A_{1}\exp\left(  -r^{2}/D^{2}\right)
\label{12}%
\end{equation}
which carries a topological charge $m=1$. The beam intensity vanishes at
$r=0$, reaches its maximum value $A_{\max}=A_{1}D\exp\left(  -0.5\right)  $ at
$r_{\max}=D$, and decays exponentially for larger $r$. Here $D$ is the
characteristic transverse width of the structure, and $A_{1}$ is a constant
which, together with $D$, determines the electromagnetic beam power $P=\frac{\pi}{4}
A_{1}^{2}D^{4}$.

We numerically integrate Eqs. (4)--(5), assuming $P<P_{c}$.
For powers below a threshold value $P<P_{0}\approx145$, the beam undergoes
diffraction. In contrast, for intermediate powers
$P_{0}<P<P_{c}$, the evolution is governed by nonlinear self-organization and
instability. When the initial beam parameters lie within the basin of
attraction of the stationary vortex soliton, and the initial amplitude is
chosen close to the soliton value $A_{\max}\approx\widehat{A}_{m}$, the
Gaussian vortex exhibits partial relaxation toward the ground-state vortex
soliton through radiation losses. This quasi-relaxed state remains azimuthally
unstable and subsequently breaks up into multiple filaments. The filaments are
expelled tangentially with respect to the initial ring-shaped intensity
distribution and evolve into stable, spatially localized solitonic beams just as it occured for stationary inital conditions. \ If the initial Gaussian beam parameters $(P,A_{m})$
lie far outside the basin of attraction of the vortex soliton, the structure
does not approach the soliton configuration. Nevertheless, the qualitative
evolution remains similar, with filamentation occurring at earlier propagation distances.

Similar behavior is observed for vortex beams with higher topological charge.
It should be noted, however, that high-charge vortices are generally believed
to be unstable even during vacuum propagation due to topological
considerations, which imply that such beams tend to decay into multiple
single-charge vortices. This decay, however, is governed by an algebraic
(secular) instability, whereas the breakup observed in our system is
exponential. Consequently, the secular instability does not have sufficient
time to develop before the onset of the nonlinear azimuthal instability. We
emphasize that optical vortices possess a topological nature, corresponding to
branch points where both the real and imaginary parts of the field vanish
simultaneously. The topological charge is defined by the number of
intersecting pairs of zero lines of the real and imaginary components of the
field. As a result, a vortex embedded in an electromagnetic beam cannot
disappear under continuous deformations of the field, even when the beam
undergoes substantial structural changes. An important and experimentally
attractive feature of the breakup process described above is that, despite the
fragmentation of the beam, the field amplitude at the center of the structure
remains strictly zero throughout the evolution. This topological robustness
provides a clear and unambiguous signature that can be exploited for
experimental verification.

Although throughout most of this manuscript we have focused on the
ultrarelativistic degeneracy regime ($d=1$) we have also performed analogous
studies for other values of the degeneracy parameter, namely $d=0.5$ and
$d=0.01$ corresponding to plasma densities $n_{0}=5.96\times10^{29}cm^{-3}$
and $n_{0}=5.96\times10^{26}cm^{-3}$, respectively. In all cases, the
underlying physics and the qualitative features of the beam dynamics remain
essentially the same as those obtained for $d=1$. Both the threshold power $P_{0}$ and the admissible field
amplitudes decrease with decreasing $d$. For example, we find $P_{0}\approx120$ for
$d=0.5$ and $P_{0}=95.$ for $d=0.01$.

Restoring the dimensions, the corresponding threshold power can be written as%

\begin{equation}
P_{0}^{\left(  \dim\right)  }\simeq0.17\left(  \frac{\omega_{0}}{\omega_{e}%
}\right)  ^{2}\Gamma_{0}R_{0}^{2}P_{0}\text{ }GW, \label{13}%
\end{equation}
where $P_{0}^{\left(  \dim\right)  }$ is measured in GW. \ For
degeneracy parameter in the range $d=0.01-0.5$ , required beam
power lies in the interval $P_{0}^{\left(  \dim\right)  }=(15.6-20.5)GW$. Since
plasma is assumed to be transparent, the photon energy of the
beam must satisfy $\hbar\omega_{0}>>(1KeV-29KeV)$, placing the relevant
frequency range in the hard $X$-ray band.

\section{Conclusions}

We have studied the propagation and stability of electromagnetic vortex beams
in relativistically degenerate plasmas using a self-consistent relativistic
fluid--Maxwell model. We demonstrated the existence of localized vortex
solitons carrying orbital angular momentum and analyzed their linear and
nonlinear stability.

Vortex solitons are shown to undergo azimuthal symmetry-breaking instabilities
whose growth rates depend on the beam power, propagation constant, and
topological charge. The dominant unstable mode determines the number of
filaments formed during breakup, in agreement with linear stability theory.
Although multiply charged vortices are known to exhibit slow, algebraic
splitting even in vacuum, the instability identified here is exponential and
develops on much shorter scales, making secular decay irrelevant in dense plasmas.

We further examined the nonlinear evolution of non-ideal Gaussian vortex beams
and found that vortex solitons act as nonlinear attractors with a finite basin
of attraction. Depending on the initial parameters, beams either diffract or
undergo partial relaxation toward the stationary vortex soliton followed by
filamentation. Throughout the evolution, the vortex core remains topologically
protected, with the field intensity at the beam center remaining strictly zero.

These results persist across a broad range of degeneracy parameters, although
the admissible power range decreases with decreasing degeneracy. Estimates of
the dimensional threshold power place the relevant regime in the hard $X$-ray
band and at multi-gigawatt beam powers, conditions naturally realized in dense
astrophysical environments near compact objects.

\bigskip

The research was supported by the Shota Rustaveli National Science Foundation
grant No. FR-24-1751. The research of N.M. was supported by the Knowledge Foundation
at the Free University of Tbilisi.

\newpage

\bigskip

\bigskip






\end{document}